\documentclass[namedreferences]{solarphysics}

\usepackage[hyperref,optionalrh]{spr-sola-addons} 
\usepackage{graphicx}        
\usepackage{color}           
\usepackage{breakurl}        
\usepackage{upgreek} 

\chardef\us=`\_

\begin{document}

\begin{article}
\begin{opening}

\title{Solar Total Eclipse of 21 August 2017: Study of the Inner Corona Dynamical Events Leading to a CME}

\author[addressref={aff1,aff2},corref,email={bfilip@izmiran.ru}]{\inits{B.}\fnm{Boris}~\lnm{Filippov}}
\author[addressref=aff2,email={koutchmy@iap.fr}]{\inits{S.}\fnm{Serge}~\lnm{Koutchmy}}
\author[addressref={aff2,aff3},email={nicolas.lefaudeux@gmail.com}]{\inits{N.}\fnm{Nicolas}~\lnm{Lefaudeux }}


\address[id=aff1]{Pushkov Institute of Terrestrial Magnetism, Ionosphere and Radio Wave Propagation, Russian Academy of Sciences, Troitsk, Moscow, 108840, Russia}
\address[id=aff2]{Institut d'Astrophysique de Paris, CNRS and Sorbonne University, UMR 7095, 98 Bis Boulevard Arago, 75014 Paris, France}
\address[id=aff3]{Imagine Eyes, 18 rue Charles de Gaulle, 91400 Orsay, France}

\runningauthor{B. Filippov  {\it et al.}}
\runningtitle{Inner Corona Dynamical Events}

\begin{abstract}
	Total solar eclipse (TSE) coronal large  and small scale events were reported in the historical literature but a definite synoptic coverage was missing for studying a relationship with the  more general magnetic context of the solar-disk. We here analyze temporal changes in the solar corona before, during, and after the total solar eclipse on 21  August 2017 from a set of ground-based  and of space-borne observations. High-quality  ground-based white-light (W-L) observations and a deep image processing allow us to reveal these changes for the first time with a fraction of a one-minute time resolution. Displacements of a number of fine coronal features were measured for the first time at these small radial distances, using a diffraction limited instrument  at a single  site. The comparison with space-based observations, including observations from the {\it Solar Terrestrial Relations Observatory} (STEREO) mission, showed that the features belong to a slow coronal mass ejection (CME) propagating through the corona with the nearly constant speed of 250\,km\,s$^{-1}$. Our TSE images provide the same typical velocity as measured at a distance of one solar radius from the surface. The event was initiated by coronal dynamics manifested by  a prominence eruption that started just before the eclipse observations and an ascent of a U-shaped structure visible in the AIA 171 \AA \ channel, which we assume as the lower part of a coronal cavity.   The prominence material was observed draining down towards the chromosphere along the prominence arch. While the prominence disappears in the STEREO-A field-of-view at the height of about 6$^{\prime}$ above the limb, the corresponding flux rope seems to continue  towards the outer corona  and produces the slow CME with turbulent motion. The overall mass of the moving features is evaluated based on absolute photometrical data extracted from our best W-L eclipse image.
\end{abstract}
\end{opening} 

\section{Introduction}
     \label{S-Introduction} 

Coronal mass ejections (CMEs) were observed for the first time in the early 1970s from space-borne observations of the {\it Orbiting Solar Observatory} (OSO)-7 coronagraph \citep{To73} and  were called ``coronal transients''.  The following generations of space-borne coronagraphs drastically enlarged the knowledge of the phenomenon. It was proved that these observed transient changes in the structure of the solar corona really represent mass motions in space \citep{Wa84,Sc96}  supported by changes near the solar surface. 

In  past eclipse observations, there was a chance to discover a large scale dynamical coronal event. If a CME were well outlined in the coronal structure and observations were made at different places far along the Moon-shadow path, the comparison of two eclipse images could provide information about moving features. However, CMEs were not really discovered before the space era \citep{Wa84}. There are structures in drawings of the corona made in the 19th century that nowadays could be considered as CMEs \citep{Bu49,Ed74,Cl89}, but they did not attract too much attention and were not recognized as moving. The best eclipse coronal event we found, on 21 September 1941,  was analyzed in the little-known \citet{Bu49} work dealing with coronal structures . The introduction of the work gives an excellent account of historical eclipse observations suggesting dynamical phenomena at a time when the true high-temperature nature of the corona was not yet fully recognized. For this well-observed 1941 eclipse event, her Figure 49  shows the proper motion map of the prominence during the event; the motion of fine white-light (W-L) coronal details (loops, front, rays) in the inner corona is also of great interest, see Figure 1a. Her Table 15 giving a large number of observed parameters, including the measured displacements from several plates separated in time by an interval up to nine minutes, convincingly demonstrates that a coronal event with complex interlaced loops and rays showing  outwards velocities of 10 to 50\,km\,s$^{-1}$, are connected to the dynamical behaviour of the underlying prominence; see Figure 1a. Rapid variations in the coronal structure were indeed recorded as early as the 1973 total solar eclipse (TSE) in images obtained with a two-hour interval \citep{Ko73}.

Another historical observation of a large eclipse CME event called  a transient event was reported from the analysis of pictures obtained 27 minutes apart and 90 minutes apart during the 15 February 1980 TSE, see Figure 1b taken from the \citet {Ai94}  report. This event was called, in the citizen science literature relating TSE observations, the ``tennis racket event''. Twisted diverging legs were suggested by \citet{Ai94}  from the visual interpretation of the processed pictures, with a possible reconnection occurring in the low part of the twisted loop as the result of a kink instability; large velocities were measured higher up, typically 400 to 800\,km\,s$^{-1}$ depending of the radial distance. Only low-resolution-photographic pictures without  space-based observations were available on 15 February 1980. However this photographic observation suggested quite convincingly that the apparent bubble-like configuration of a CME does not hold in 3D because a twisted flux rope was identified at large scale.  

 \begin{figure}   
   \centerline{\includegraphics[width=1.\textwidth,clip=]{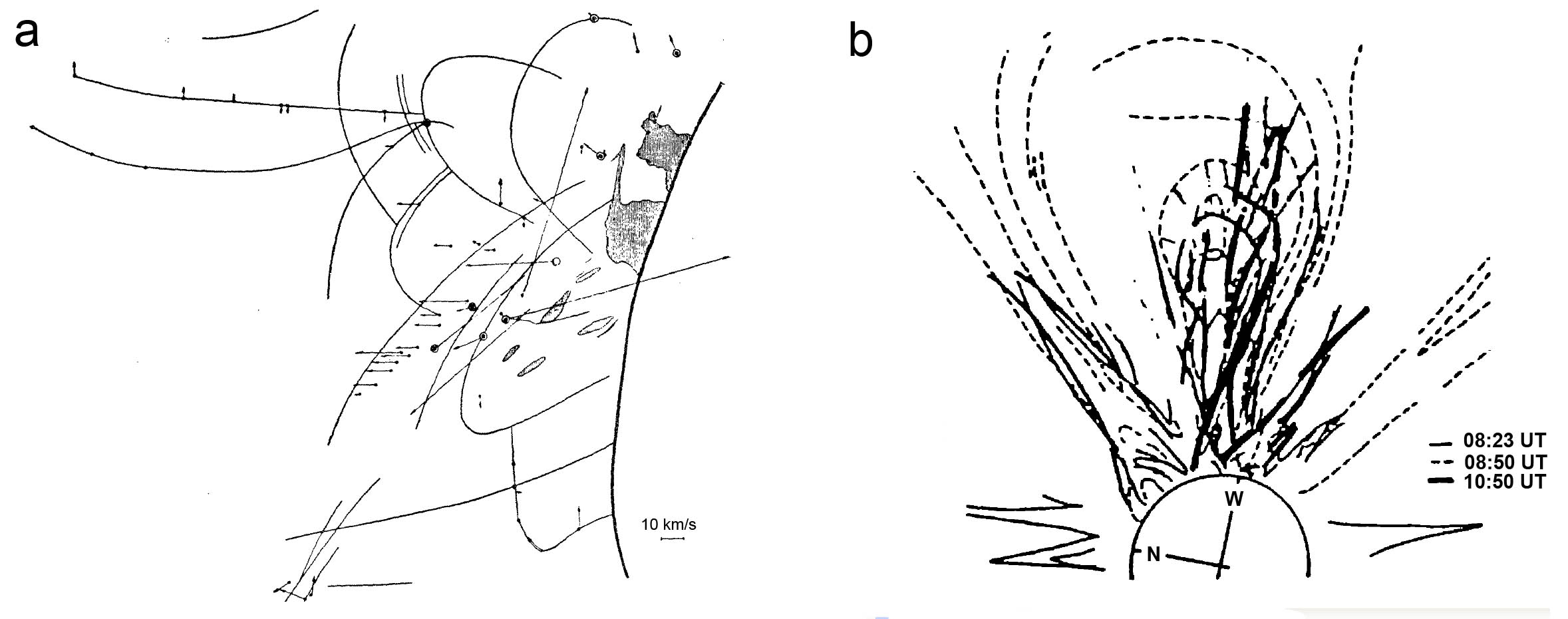}              }
              \caption{{\it Left:} Historical observation of the TSE coronal event of 21 September 1941 with slow expansion of coronal loops and rays above an activated prominence. The bar at the bottom right corresponds to a proper motion of 10\,km\,s$^{-1}$. The prominence dynamics were discussed separately but one can note the prominence remnants of cool material just under a W-L front with expansion suggesting a synchronized eruption. Eclipse multi-station observations and analysis by \citet{Bu49}. {\it Right:} The so-called ``tennis racket'' CME  event observed during the  15 February 1980  TSE; drawing made from large scale pictures (\citet{Ai94}, Figure 2B, with permission of the authors).                    }
   \label{F1}
   \end{figure}

More recently, an eclipse CME was well described by \citet{Ko04}. It occurred during the  11 August 1999  widely observed eclipse corona near the maximum of activity. At the W-limb a large prominence cavity system evolves; rather large velocities (440\,km\,s$^{-1}$) were reported from the  Large Angle and Spectrometer Coronagraph (LASCO: \citealp{Br95}) C2  W-L observations onboard the {\it Solar and Heliospheric Observatory} (SOHO). The authors developed a great deal of attention to the phenomenon of expansion observed in the high arch overlying the erupting part of the prominence without reaching a definite conclusion about the origin of the large CME, although some evidence of a reconnecting phenomenon was given based on a sequence of  the  processed filtergrams taken by the Extreme-ultraviolet Imaging Telescope (EIT: \citealp{De95}) onboard SOHO. 

At the time of the following solar maximum, \citet{Ha14} analyzed again some eclipse CMEs seen at both limbs of the eclipse corona of  13 November 2012, using two sets of images taken 35 minutes apart. At the E-limb a rather faint bubble-like CME related to a flare was studied showing a velocity of the front of 360\,km\,s$^{-1}$ in the region one solar radius (R$_\odot$) from the limb, which is low for a solar maximum CME. At this well-observed eclipse, many observations were taken from Australia, but unfortunately the CMEs were not noticed although they were well recorded afterwards with the LASCO coronagraphs. At the W-limb the unique eclipse pictures of  \citet{Ha14} show the motion of a system of arches that the authors identified with a conventional CME case appearing after a prominence eruption with a dark cavity moving outwardly. The proper motions were typically 10\,km\,s$^{-1}$ in the inner corona region and 130\,km\,s$^{-1}$ in the 1\,R$_\odot$  region. They called the event a ``hot-air balloon loop structure'' and proposed to identify it with the initiation region of the CME as shown by the eclipse pictures. The corresponding  SOHO/LASCO CME was seen with velocities of 130\,km\,s$^{-1}$ at the height of 4\,--\,5 R$_\odot$  which makes it definitely a slow CME observed at solar maximum.  Similar ``balloon-like'' structures were observed previously ({\it e.g.}  \citealp{Sr99,Sr00}).

During the TSE of 2013, two large prominent CMEs extending far into the corona can be seen in the coronal image of  03  November 2013 (\textsf{www.zam.fme.\\vutbr.cz/$\sim$druck/Eclipse/Ecl2013g/TSE\_2013c2\_ed/Hr/TSE\_2013c2\_ed.png}). The eclipse reconstructed image is a result of a sophisticated digital processing of many coronal images taken with different exposures at the observation site using the M. Druckm\"uller's method \citep{Dr06} . The CMEs can be easily recognized in the processed image and compared with space-based coronagraphic observations. However, the authors did not use a method that would allow detecting a CME motion or even small changes in the structure of the corona during the short time of totality at a particular observation site from the same instrument. No space-based observations of the same CME were analyzed  as far as we know. 

The CME speed in the sky plane varies in a wide range from tens to more than 2500\,km\,s$^{-1}$, with an average value of about 500\,km\,s$^{-1}$ \citep{Go04}. CMEs with  lower than the average speed can be considered as slow \citep{Bo97}, while the others are fast. \citet{Sh99}  suggested that slow, or as they call them gradual CMEs, accelerate within the coronagraph field-of-view (FOV) and in most cases are associated with filament eruptions. Slow CMEs with persistently weak acceleration are known also as balloon-type events  \citep{Sr99, Sr00}. Fast, or impulsive, CMEs, in contrast, usually decelerate and are related to solar flares. However, the separation of CMEs into two groups is rather arbitrary because parameters of flare-associated and non-flare CMEs considerably overlap  \citep{Vr05} and in most energetic events both flares and filament eruptions are observed ({\it e.g.} \citealp{Sc15}). Many researchers agree that one mechanism is sufficient to explain flare-related and prominence-related CMEs  \citep{Ch03,Fe04}. 

The typical morphological structure of a CME consists of three parts: a bright frontal loop followed by a dark cavity with an embedded bright core representing remnants of an eruptive filament  \citep{Ho81,Il85}. The CME dark cavities result from the development of coronal cavities observed in W-L, extreme ultraviolet (EUV), and soft X-ray emission around prominences and at the feet (bases) of coronal streamers \citep{Gi06,Ha10,Ku12,Re12,Ka15}. It is widely assumed that the cavity represents a magnetic flux rope with helical magnetic field. The lowest sections of the helices, dips, can be filled with colder dense plasma of a prominence. Prominence material is sometimes observed as highly twisted structures in the core of CMEs \citep{Bo97,Li15}.

A CME cannot be characterized by a unique  speed. Different parts of the CME move with different speeds. The fastest part of the CME is the frontal loop; the core moves slower \citep{Zh01,Fo03,Ch14}. The bottom edge of the cavity moves with the lowest speed \citep{Vr04}. This is the consequence of the composition  of the flux rope in two motions: a translational ascending motion and a radial expansion. Finally, in the framework of the study of slow and ``weak" CMEs, let us note the analysis of events that are a component of the coronal dynamics possibly related to the so-called slow wind. Such components are now called ``blobs"  \citep{Sh07}.

 In this article, we present the results of a specially designed imaging W-L experiment performed during the TSE observations on 21 August 2017. Definite traces of the propagation of a slow CME inside the  W-L corona are reported during the totality. Multi-wavelength observations with space-based instruments are also used to study the evolution of the coronal region under interest before, during, and after the totality. The CME was initiated by a very slow filament eruption and moved later with a constant speed.

\section{Observations}
\label{S-O}

W-L eclipse observations we used were made on the west coast of the United States near Unity, Oregon near 17:24 UT. A special algorithm to make high quality High Dynamic Range (HDR) composite images wos used.  The High dynamic range (HDR) imaging methods have been developed for a long time  (see, {\it e.g.},  \citealp{Ba18}). For processing eclipse coronal images the most popular algorithm was developed by Milo Druckm\"uller in Brno  \citep{Dr06} (see also his website \textsf{www.zam.fme.vutbr.cz/$\sim$druck/\\Eclipse/}). Excellent reproductions of images are provided in the website at \textsf{hdr-astrophotography.com/} as well as a view of the equipment used.   Some selected original images are available from the site: \textsf{bass2000.obspm.fr/piwigo/index.php?/\\category/141.} In addition,  EUV coronal images taken by the Atmospheric Imaging Assembly (AIA: \citealp{Le12}) onboard the {\it Solar Dynamic Observatory} (SDO) and the Sun Earth Connection Coronal and Heliospheric Investigation (SECCHI) Extreme Ultraviolet Imager (EUVI:  \citealp{Wu04,Ho08})  onboard the  {\it Solar Terrestrial Relations Observatory} (STEREO) are used. W-L images obtained with the SOHO/LASCO  were analysed for the CME evolution in the more outer corona.
 
 \begin{figure}   
   \centerline{\includegraphics[width=1.\textwidth,clip=]{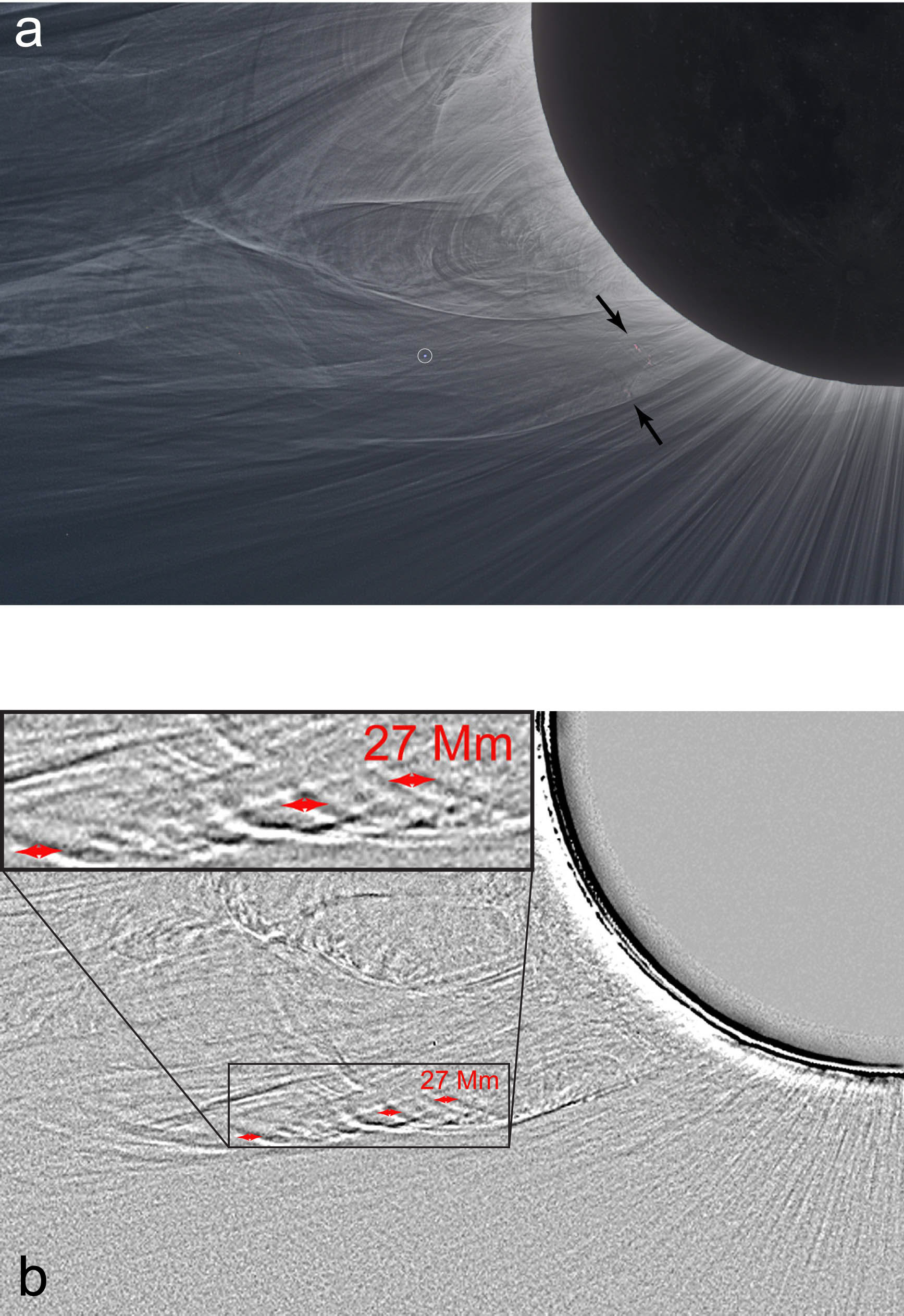}              }
              \caption{Partial frame of the ground-based coronal image on 21 August  2017  at 17:24 UT made using a deep-processing technique to show fine coronal structures {\bf (a)}, difference image of the corona with a time lapse of 1.75 minutes {\bf (b)}. {\it Black arrows} in {\bf (a)} point to the eruptive prominence remnants observed as reddish features owing to H$\alpha$ and D$_3$ emissions. {\it Red arrows} in {\bf (b)} indicate the displacement of coronal structures during the totality. The small white dot surrounded by the {\it thin white circle} near the centre of the upper frame is the image of a star not seen after making the subtraction as seen in the lower frame, in contrast to the lunar limb showing a large displacement (see also the Electronic Supplementary Material, Movie 1).                      }
   \label{F2}
   \end{figure}

The high-quality W-L coronal images allowed us to recognize a number of small coronal features moving outward in the south-east sector of the corona (Movie1) during the short period of the totality ($\approx$ 1 minute 45 seconds).  Frames in Movie1 run forth and back for better visibility of the moving features.  They are hardly noticeable in a particular static coronal image (Figure 2a) and it is unlikely that their changes  could be identified in images taken at different places along the Moon-shadow path. Only an examination of a movie  or a difference of images taken at the beginning and at the end of the totality (Figure 2b) using the same instrumental set-up and the same image processing allow us to find the changes in the fine coronal structure. The displacement of the best identified features in the vicinity of radial distances $\approx 1.8\,$R$_\odot$  is about 27 Mm as is shown in Figure 2b by red arrows. The speed of the feature motion is about 250\,km\,s$^{-1}$. In order to find the cause and source of this motion, we analyzed observations of the Sun in different wavelengths made by different space missions. 
 
\section{Prominence Dynamics} 
      \label{S-P}      

\begin{figure}   
   \centerline{\includegraphics[width=1.\textwidth,clip=]{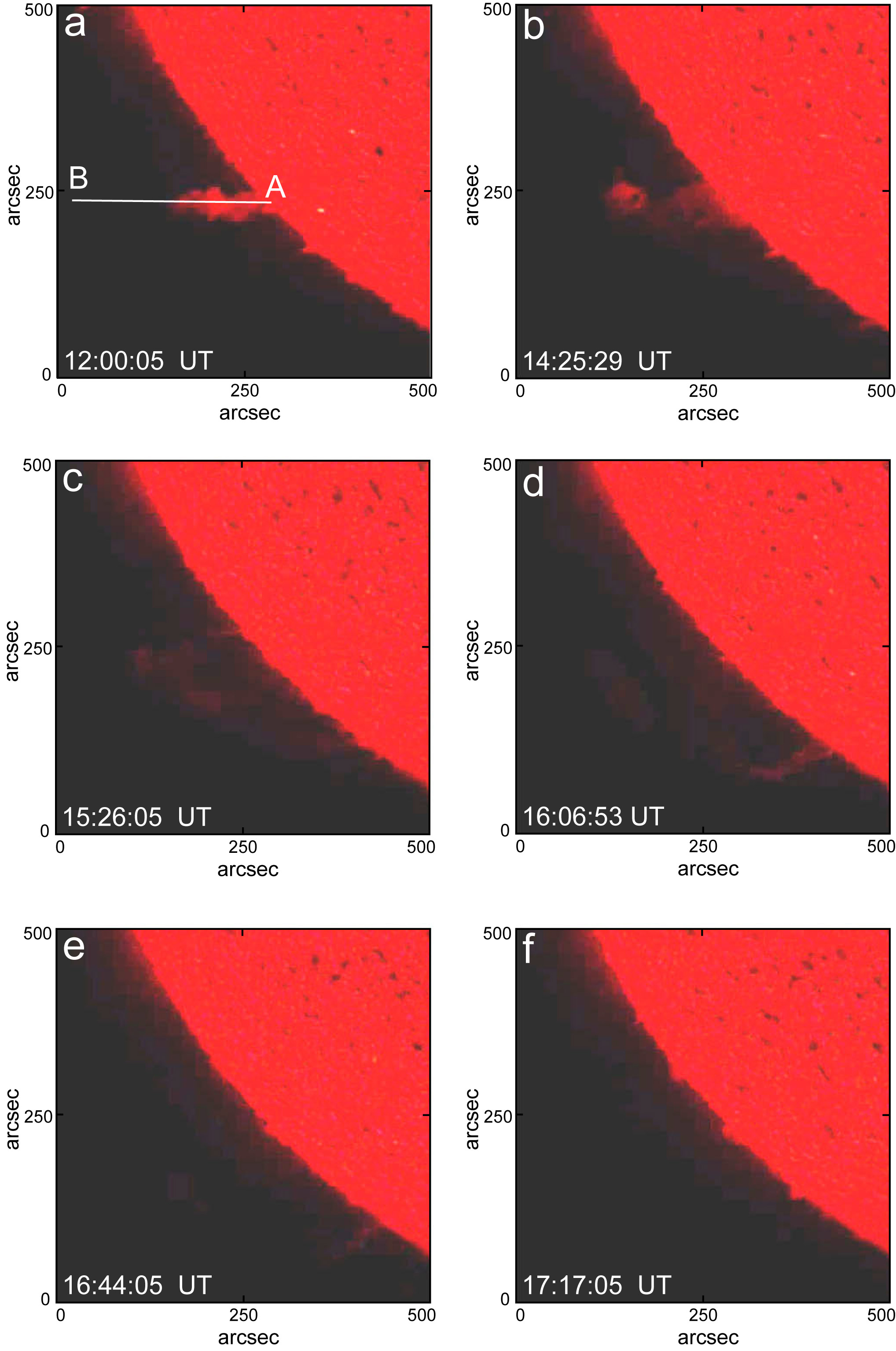}              }
              \caption{SDO/AIA 304 \AA \ sequence showing the prominence eruption of 21 August 2017. At the time of totality of the TSE (17:24 UT), the prominence is barely seen. The {\it line AB} shows the slit position for the distance--time-diagram presented in Figure 4.  (Courtesy of the SDO/AIA science team)                       }
   \label{F3}
   \end{figure}

\begin{figure}   
   \centerline{\includegraphics[width=0.8\textwidth,clip=]{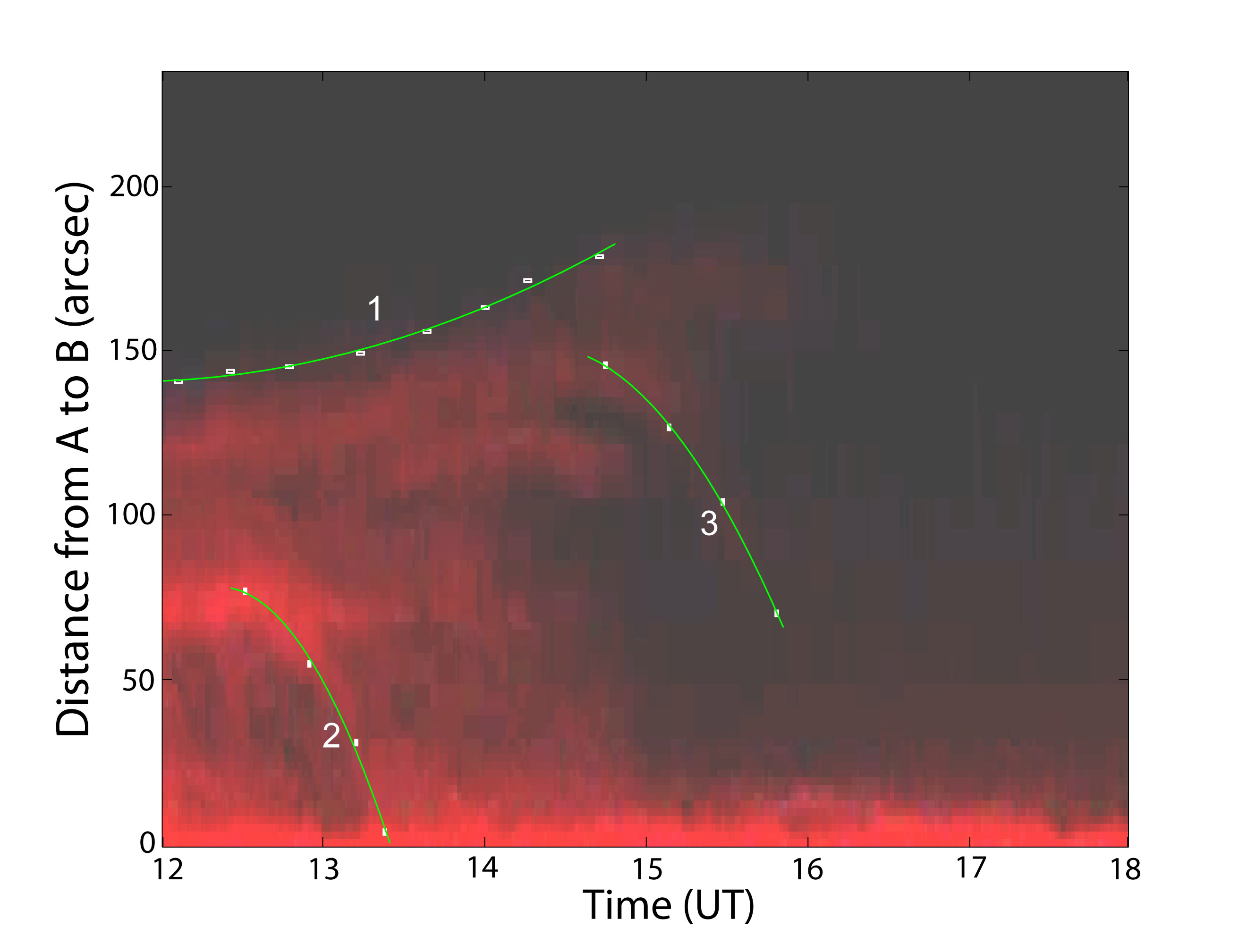}              }
              \caption{Distance--time diagram along the slit AB shown in Figure 3a.   {\it White  rectangles} show some selected points in the map. Time of the TSE is 17:24 UT. {\it Green lines} present  second order polynomial fits to the points.                }
   \label{F4}
   \end{figure}

\begin{figure}   
   \centerline{\includegraphics[width=1.\textwidth,clip=]{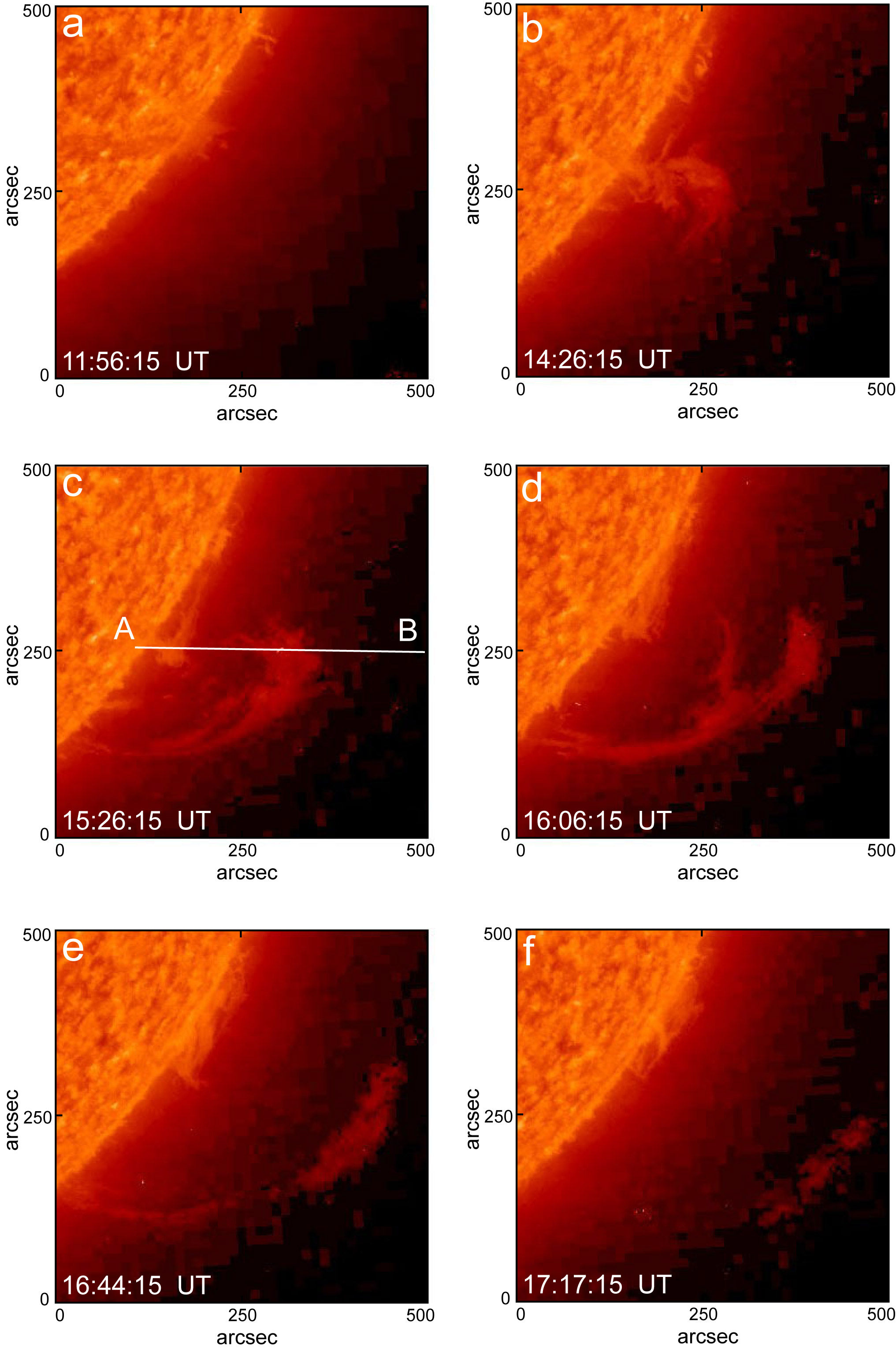}              }
              \caption{STEREO-A/EUVI 304 \AA \ images showing the prominence eruption on 21 August 2017. The  {\it line AB} shows the slit position for the distance--time-diagram presented in Figure 6.   (Courtesy of the STEREO-A/EUVI science team)                      }
   \label{F5}
   \end{figure}

\begin{figure}   
   \centerline{\includegraphics[width=0.8\textwidth,clip=]{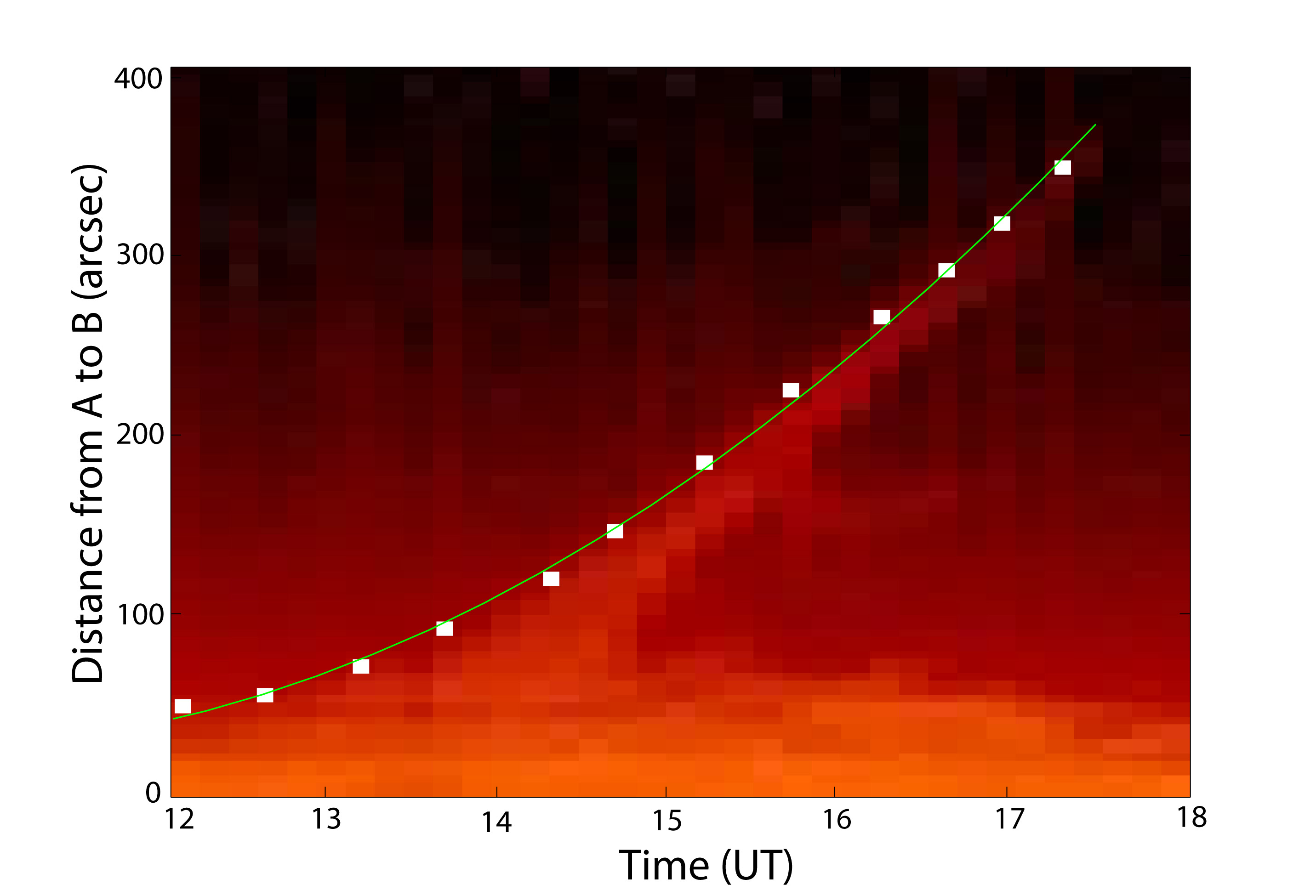}              }
              \caption{Distance--time diagram along the slit AB shown in Figure 5c.                      }
   \label{F6}
   \end{figure}

\begin{figure}   
   \centerline{\includegraphics[width=1.\textwidth,clip=]{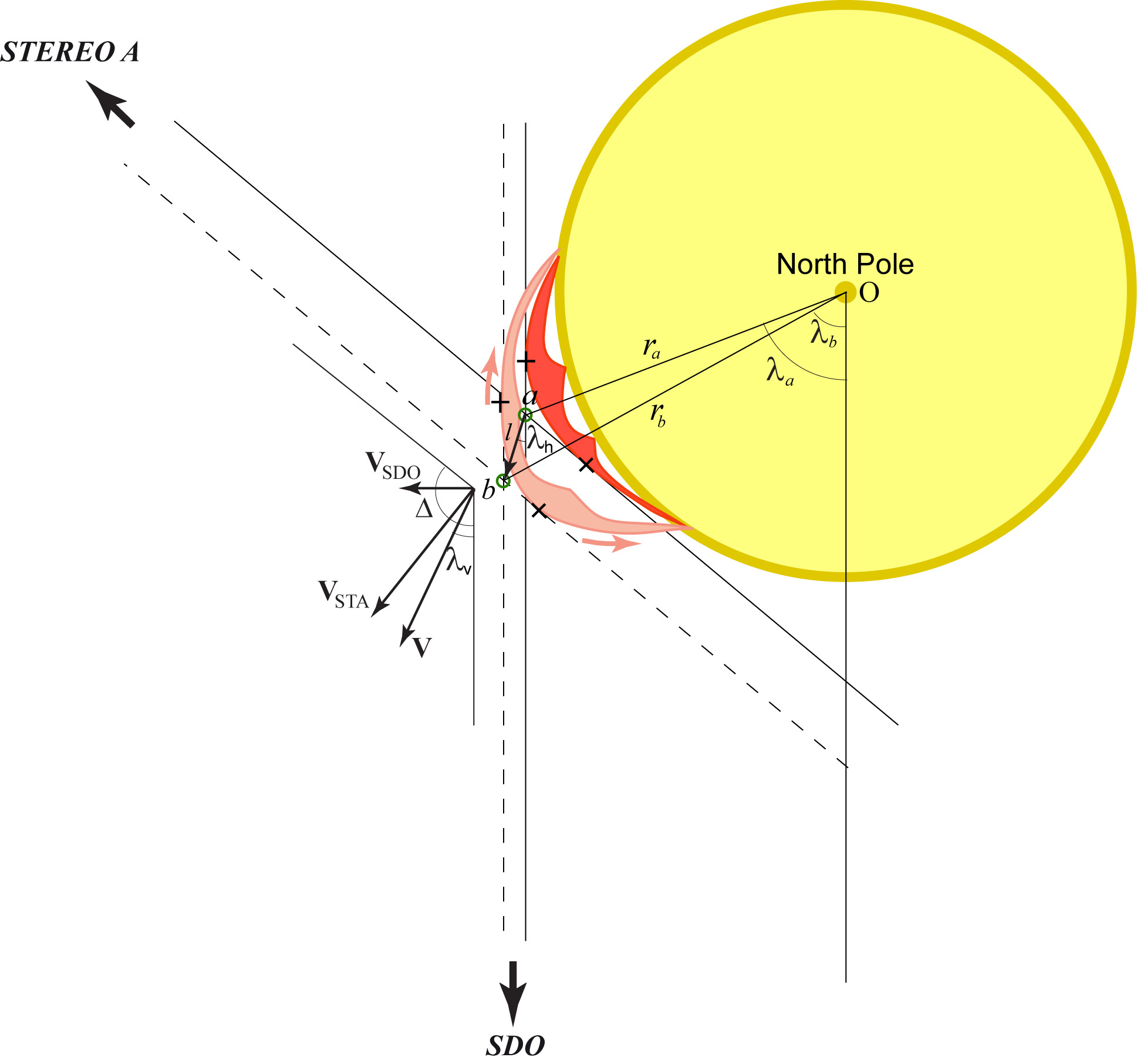}              }
              \caption{Schematic of the possible position of the prominence relative to the viewpoints of  SDO   and    STEREO spacecrafts.   The {\it yellow  circle} represents the section of the Sun at the latitude of --40$^\circ$.  {\it Red and  pink arches} show the prominence in the initial state and during the eruption, respectively.    }
   \label{F7}
   \end{figure}

Several hours before the totality observed at 17:24 UT, a prominence activation started in the south-east sector of the solar limb. It was observed in the 304\,\AA\ channel of SDO/AIA (Figures 3 and 4, Movie 2) but was more impressively visible in the 304 \AA \ channel of STEREO/EUVI (Figures 5 and 6, Movie 3). STEREO Ahead (STA) was at the 128$^\circ$ separation angle with Earth, and the prominence was in the south-west sector of the limb in its images. The prominence belonged to a family of polar crown prominences  elongated along the heliograpic parallel (East-West direction) at a latitude of about S\,40$^\circ$. The top of the prominence shifted most   notieably both in the SDO FOV and STA FOV in the direction parallel to the Equator. That is why we will analyse  the motion in  this direction  and will refer below to the distance from the prominence top to the limb along this direction as the projected height, if it is not stated otherwise. At the beginning of the activation  at 13:00 UT  the projected prominence height above the limb was about 100\,Mm  in the AIA FOV and about 25\,Mm in the EUVI FOV. If we observe the same point of the prominence spine with each instrument, we can estimate its latitude using the expression for the projected height of a feature above the limb at the latitude $\upphi$ ({\it e.g.}  \citealp{Fi13})

\begin{equation}
h = (H + \mathrm{R}_\odot \cos \upphi) \sin \lambda - \mathrm{R} _\odot \cos \upphi \:,
\end{equation}

\noindent where $h$ is the projected height of the feature above the limb, $H$ is the true distance from the feature to the surface along the line parallel to the equator, and $\lambda$ is its longitude, the angle between the feature and  the line-of-sight (LOS). Applying Equation 1 to the data obtained from two viewpoints marked by the indices 1 and 2 and taking into account that $\upphi_1 = \upphi_2= \upphi$, $H_1 = H_2= H$, $\lambda_1 +\lambda_2= \Delta$, where  $\Delta$ is the separation angle between the two viewpoints, we find from these two equations

\begin{equation}
\cot \lambda_1  = \left( \frac{h_2+\mathrm{R} _\odot \cos \upphi}{h_1+\mathrm{R} _\odot \cos \upphi} +\cos \Delta \right) \frac{1}{\sin \Delta} \:.
\end{equation}

Since the LOS for both instruments was directed along the prominence axis, the visible prominence tops did not  necessarily show the same point of the prominence spine. Figure 7 illustrates the position of the instruments relative to the prominence. Thin solid lines represent LOSs touching the prominence spine at 12:00 UT, while dashed lines correspond to the moment 15:30 UT when the prominence was erupting. If the lines touch the spine at the same point marked by the small green circle ``a'', its longitude is  $\lambda_a  = 72^\circ$. However, they can touch the spine at different points as it is shown in Figure 7. The intersections of the SDO LOS with the spine are marked by crosses, while the same for STA are marked by Xs. When the prominence axis has low curvature, the intersection points are  widely separated. 

The prominence started to rise very slowly with a low acceleration as is seen in distance--time diagrams in Figures 4 and 6 obtained along the slits shown in Figures 3a and  5c. Green lines show second-order polynomial fits to points marking the top of the prominence 

\begin{equation}
d  = a + b t + c t^2 \:.
\end{equation}
Parameters $a, b$, and $c$ are presented in Table 1 with the standard deviations for $d$ in km above the limb and $t$ in seconds. Acceleration is a little bit higher in the STA FOV but does not exceed 0.8 $\pm$ 0.1\,m\,s$^{-2}$. The prominence is visible longer for STA, and the final speed at 17:30 UT is $v_{\rm  STA}$ = 20 $\pm$ 2\,km\,s$^{-1}$ at the projected height of about $h_{\rm STA}$ = 250\,Mm. In the SDO FOV, the filament fades out after 15:30 UT at the projected height $h_{\rm SDO}$ = 140\,Mm. The  final speed  is $v_{\rm SDO}$ = 7 $\pm$ 2\,km\,s$^{-1}$. At that time $v_{\rm STA}$ = 15 $\pm$ 2\,km\,s$^{-1}$ and $h_{\rm STA}$ = 140\,Mm. It should be noted that these velocities are obtained on the basis of the distance-time diagrams with the slits parallel to the Equator.  Speeds and accelerations in the radial direction can be calculated by the multiplication by $\cos \phi$.

The highest intersections of the LOSs at 15:30 UT is marked in  Figure 7 by the small green circle ``b''. The corresponding longitude is $\lambda_b  = 64^\circ$. The displacement $l$ from the point ``a'' to the point ``b'' can be found from the triangle aOb (Figure 7) using the cosine rule 

\begin{equation}
l  = \sqrt{r_a^2  + r_b^2 - 2 r_a r_b \cos(\lambda_a - \lambda_b)} \:,
\end{equation}
where $r_a = H_a + \mathrm{R}_\odot$, $r_b = H_b + \mathrm{R}_\odot$, and $H_a$ and $H_b$ are the true (radial) heights of the points ``a'' and  ``b''. Then, from the same triangle and the sine rule  we derive the direction of displacement defined by the angle $\lambda_h$

\begin{equation}
\lambda_h  = \lambda_b - 180^\circ + \arcsin \frac{r_b \sin(\lambda_a - \lambda_b)}{l} \:.
\end{equation}
Using the values presented above we obtain  $\lambda_h  =  28^\circ$. 

On the other hand, we have the velocities of the prominence top derived from the displacement--time plots. If they correspond to projections of the true speed on the sky-planes of each instrument, we can find from geometry presented in Figure 7 the direction of the true motion from expression  analogous to Equation 2:

\begin{equation}
\cot \lambda_v  = \left( \frac{v_{\rm STA}}{v_{\rm SDO}} +\cos \Delta \right) \frac{1}{\sin \Delta} \:,
\end{equation}
which gives us $\lambda_v  =  27^\circ$. Thus,  the two estimated directions of motion are nearly the same, which shows that  the prominence spine points recorded as the tops by the instruments from the different viewpoints are close to each other.

No prominence plasma moving upward is visible either in SDO/AIA or STE\-REO/EUVI observations after 17:30 UT. The prominence disappears in the STA FOV at the height of about 6$^{\prime}$ (Figures 5f and 6). Remnants of the eruptive prominence are recognized in coronal W-L images as small reddish features having this colour evidently due to H$\alpha$ and D$_3$ emissions. They are indicated by black arrows in Figure 2a. The prominence remnants do not show any measurable displacement during the totality, as seen in Movie 1 and Figure 2b. However,  the projected radial height of the  lower reddish remnant of the filament is 115\,Mm, which is nearly  the same as the height of the last recognizable fragment of the prominence in the SDO FOV; the height of another remnant is 200\,Mm. It belongs to the upper parts of the prominence, and hence the displacement from 16 UT corresponds to a radial speed of 20\,km\,s$^{-1}$. 

During the slow ascending motion of the prominence, its material is observed to drain down to the surface. Most of the material is drained down along the southern leg of the visible arch. Accordingly, the supposed geometry (Figure 7) shows that the material falls down to the western end of the prominence arch, which is at the same time more southward. In SDO/AIA images, the material drops on the disc along the western leg causing some brightenings on the surface, while in STEREO/EUVI images it disappears behind the limb. In Figure 4, several descending trajectories can be recognised. The two most prominent trajectories marked as curve 2 and curve 3 show a projected acceleration of 15 $\pm$ 5\,m\,s$^{-2}$ and 8 $\pm$ 1\,m\,s$^{-2}$ , respectively. The projected speed at the lower parts of each curves is about 35\, km\,s$^{-1}$. The downward acceleration is more than an order of magnitude higher than the upward acceleration of  the erupting prominence spine. On the other hand, the downward acceleration is more than an order of magnitude lower than the free-fall acceleration ($\approx$ 270\,m\,s$^{-2}$ ), which shows that blobs fall down along gently sloping magnetic field lines.

Flow speeds in the leg are similar in SDO and STA FOVs. The most prominent blob near
the limb in the STA FOV is observed within the time interval 16:24\,--\,16:44 UT with the
average speed of about 100\,km\,s$^{-1}$. The most prominent blob near the limb in the SDO FOV
is observed within the time interval 16:09 - 16:19 UT with the average speed of about 70\,km\,s$^{-1}$. It is possibly the same blob but at a greater height according to the suggested
geometry (Figure 7) and it is moving at this time along a more gently sloping trajectory.

\begin{table*}
\caption{Parameters of the second-order polynomial fits to the height--time curves of Figures 4, 6, and 9}
\label{T1}
\begin{tabular}{@{}llll}
\hline
Curve & $a \; (10^3)$ & $b$ & $c \; (10^{-3})$
\\
\hline
SDO/AIA 304 \AA \ - 1 &  \, 99 $\pm$ 1  & \, 7.7 $\pm$ 0.6 & \,0.26 $\pm$ 0.05 \\
SDO/AIA 304 \AA \ - 2 &  \, 51 $\pm$ 2.5  & --5.9 $\pm$ 5.4 &\,--7.4 $\pm$  2.5 \\
SDO/AIA 304 \AA \ - 3 & 104 $\pm$ 1  &  --9.0 $\pm$ 1.5 & \,--3.9 $\pm$  0.5  \\
STA/EUVI 304 \AA \ & \, 17 $\pm$ 3 & \, 4.7 $\pm$ 0.7 & \,0.41 $\pm$  0.04  \\
SDO/AIA 171 \AA \ &  120 $\pm$ 1 & \, 3.1 $\pm$ 0.3 & \,0.19 $\pm$  0.024 \\
\hline
\end{tabular}
\end{table*}

\section{Coronal-loop Motion and the Slow CME} 
  \label{S-C}

\begin{figure}   
   \centerline{\includegraphics[width=1.\textwidth,clip=]{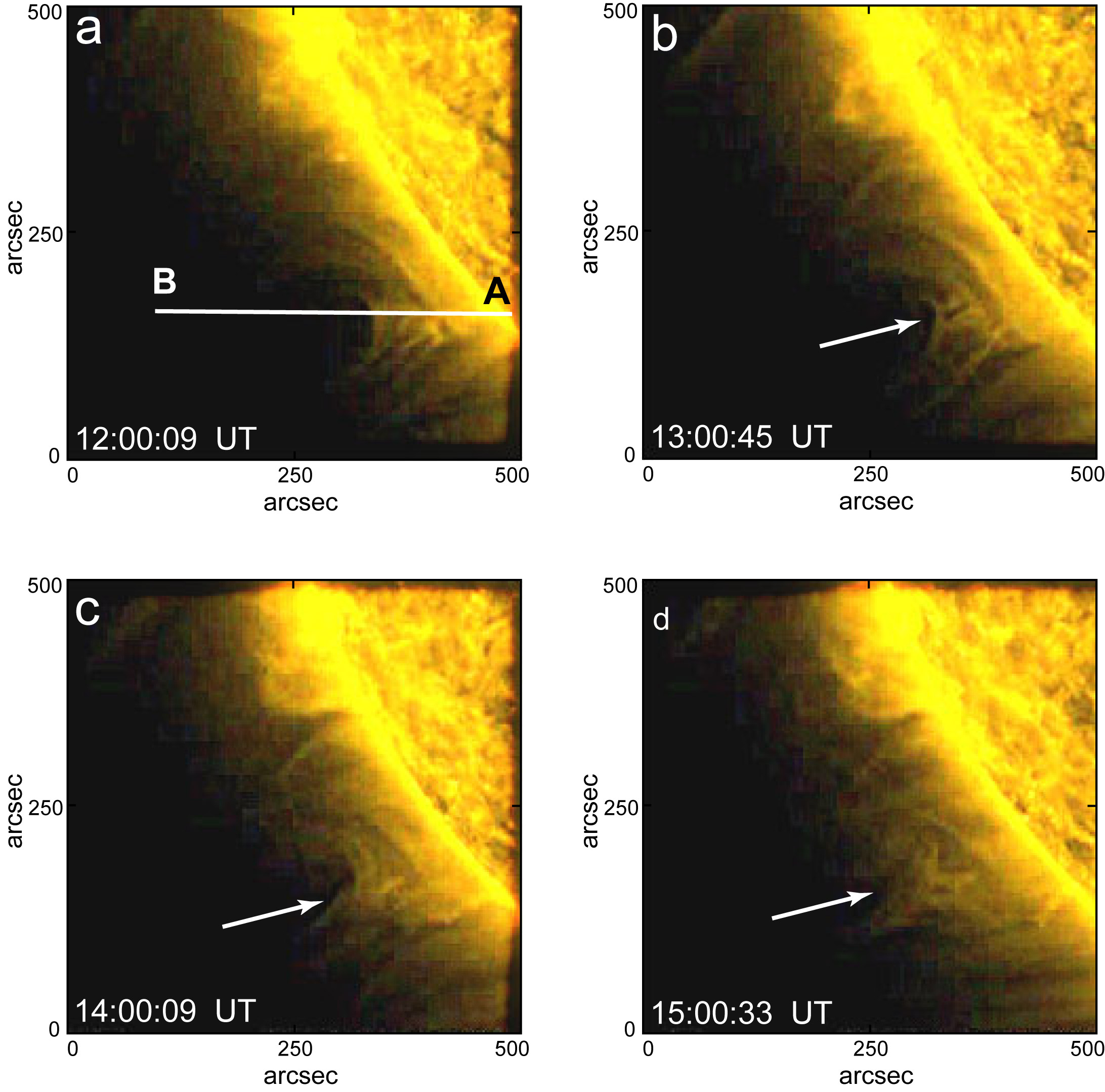}              }
              \caption{SDO/AIA 171\,\AA \ images showing the rising of the dark cavity and the coronal loops on 21 August 2017. The  {\it line AB} shows the slit position for the distance--time-diagram presented in Figure 9. (Courtesy of the SDO/AIA science team)                       }
   \label{F8}
   \end{figure}

\begin{figure}   
   \centerline{\includegraphics[width=0.8\textwidth,clip=]{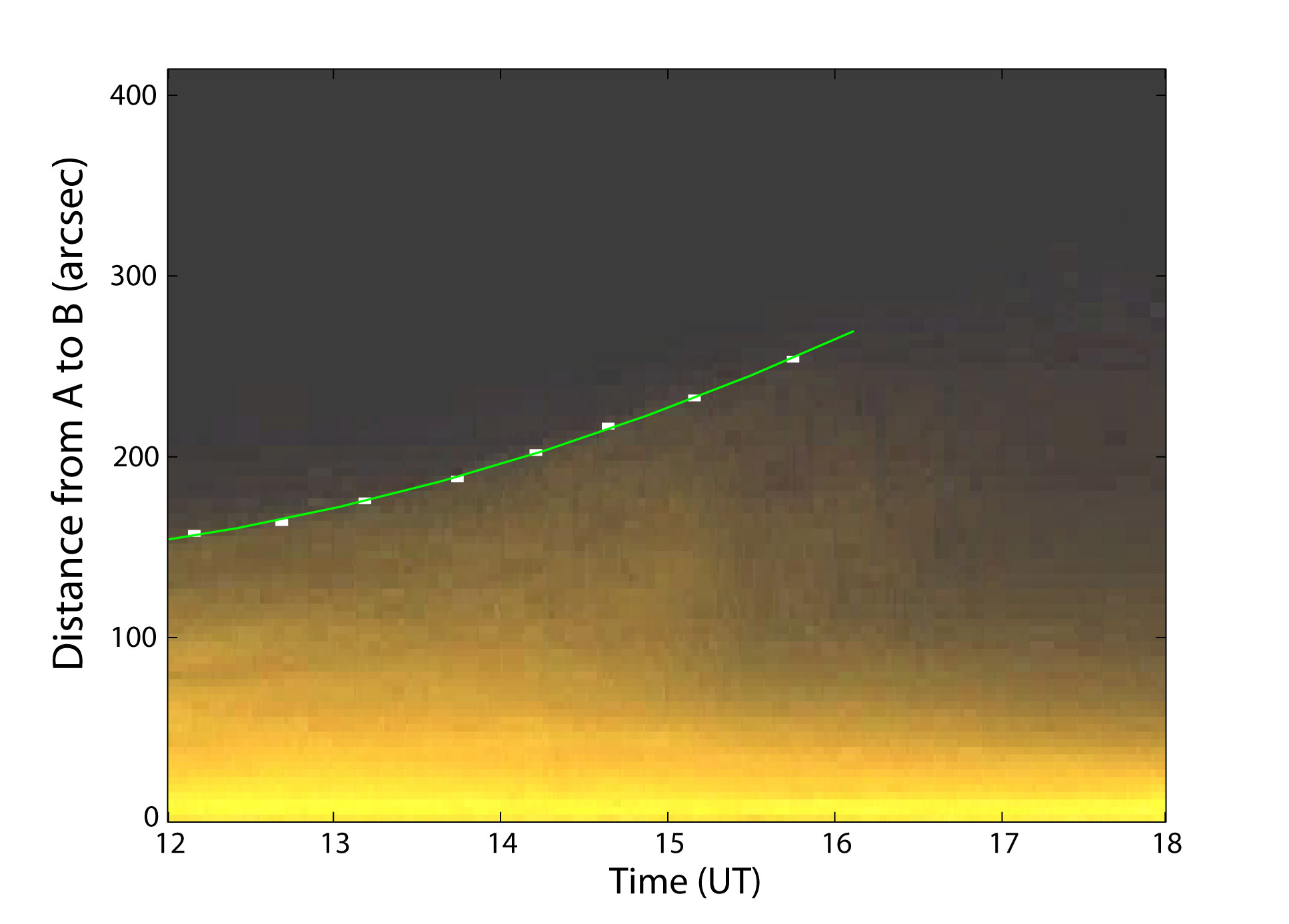}     }
              \caption{Distance--time diagram along the slit AB shown in Figure 8a.      }
   \label{F9}
   \end{figure}

In the EUV coronal channels showing a hotter  plasma than what is visible in the transition-region line of the 304 \AA\ channel, the ascending motion of coronal structures started also at about 12:00 UT (Figures 8 and 9, Movie 4). Since the polar crown prominences are associated with magnetic-flux ropes stretched in the East-West direction by the differential rotation, the flux rope axis is close to the LOS at the limb. This is a favourable position to observe coronal cavity surrounding a prominence; it could represent the internal part of the flux rope \citep{Gi06,Ha10,Ku12,Re12,Ka15}. The prominence top is located somewhere in the middle of the cavity. The U-shaped feature marked by the white arrow in Figure 8 presumably represents the lower edge of the cavity. The structure is very similar to the ones reported by \citet{Re11} and \citet{Su15}. Different parts of a cavity usually move with different speeds during an eruption, since the cavity ascends as a whole and expands \citep{Lo18}. Thus, the lower part moves with the lowest speed, while the upper part rises with the highest speed. The projected speed of the U-shaped feature is a little lower than 10\,km\,s$^{-1}$ (Figure 9, Table 1), which is similar to that of the prominence top (Figure 4). The upper part of the cavity is not visible in EUV images. While the cold prominence arch seems to disappear mostly due to the draining of the material toward the chromosphere, the hotter coronal structures fade away close to the upper boundary of the SDO/AIA FOV. There are no data about the evolution of the structure in the interval between the SDO/AIA FOV and the edge of the occulting disc of the SOHO/LASCO-C2 except for the moment of totality at Unity around 17:24 UT. 

\begin{figure}   
   \centerline{\includegraphics[width=1.\textwidth,clip=]{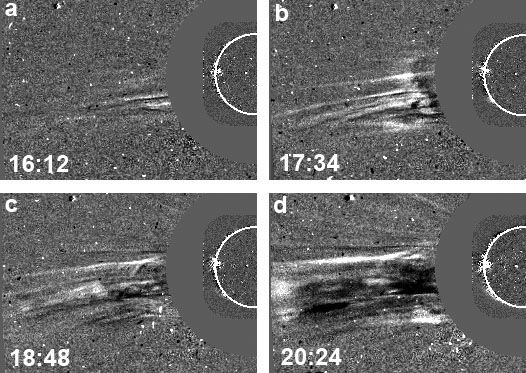}              }
              \caption{SOHO/LASCO-C2 difference images on 21 August 2017 with SDO/AIA 193 \AA \ difference images placed at the location of the disk of the Sun. (Courtesy of the SOHO/LASCO and SDO/AIA science teams)                      }
   \label{F10}
   \end{figure}

\begin{figure}   
   \centerline{\includegraphics[width=0.5\textwidth,clip=]{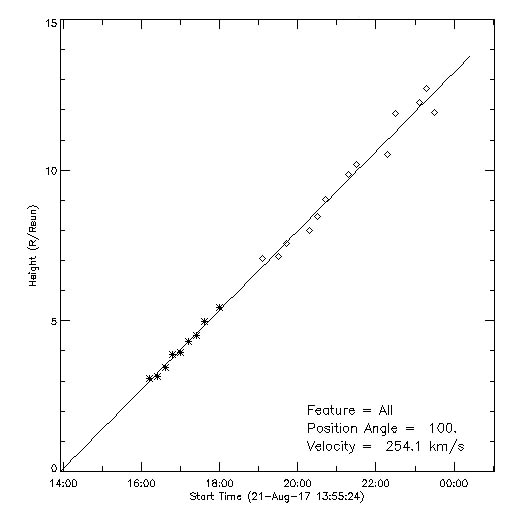}              }
              \caption{Linear fit to SOHO/LASCO observations of the E-limb CME on 21 August 2017 from the SOHO/LASCO CME Catalog (\textsf{cdaw.gsfc.nasa.gov/CME\_list/}). Velocities taken from the Brussels CACTus catalog (see \textsf{sidc.oma.be/cactus/}) give a large dispersion around the same typical value given here.                       }
   \label{F11}
   \end{figure}

According to the SOHO/LASCO CME Catalog \textsf{cdaw.gsfc.nasa.gov/CME\_list/}, a CME associated with this event appeared in the FOV of the C2 coronagraph at 16:12 UT (Figure 10). The CME moved in the SOHO/LASCO FOV from 3 to 15 R$_\odot$  with a nearly constant speed of 250\,km\,s$^{-1}$ (Figure 11). It had  neither a prominent frontal loop nor a bright core. The start time of the CME is estimated in the Catalog as 14 UT, which is close to the conspicuous  prominence activation. We assume this CME to be the continuation of the slow prominence eruption. Although there are no continuous observations of the event from the low corona to the coronagraph FOV, the W-L observations fill the gap between the SDO/AIA and SOHO/LASCO observations. The W-L eclipse images (Figure 2, Movie 1) show several coronal features moving with a speed of about 250\,km\,s$^{-1}$ at distances of 0.8\,--\,1.3\,R$_\odot$ . It suggests that the flux rope has accelerated after leaving the SDO/AIA FOV. However, in the SDO/AIA images the lower parts of the flux rope with the U-shape are visible, while in the W-L images the curvature of the features is directed downward, which implies that they belong to the upper parts of the flux rope or that they are pushed away by the surrounding magnetic loops and evolve synchronously  as suggested by the Electronic Supplementary Material movie. As was mentioned, the lower part of an erupting flux rope always moves with a lower speed, while the upper part rises with a higher speed. 

\section{Evaluation of the Density Excess and  of the Mass of the CME}

Using the absolute photometry of the well-exposed W-L image, it is possible to compute the electron densities corresponding to the intensities in excess of the CME identified in Figure 2 and well observed  when crossing the FOV of the LASCO-C2 coronagraph (see Figure 10).  The method used to deduce electron densities and the needed  formula are given in  textbooks ({\it e.g.}  \citealp{Sh65, Zi66, Bi66}) and an extended discussion of the inversion question to deduce densities, including the needed formula  were given by \citet{No96}. The important parameters that are needed to extract from the coronal image are the absolute values of intensity excess measured over the background corona along the features of the CME defined as a definitely moving set of elongated features (blobs or clouds or fragments of loops) (see Figure 2). Recently such evaluation was performed for the plasma cloud corresponding to a failed small scale ejection observed at the TSE of 2010  \citep{Ta18}  at a radial distance near 1.7\,R$_\odot$.  A more general and simple presentation is given by \citet{Ko94}. 

We selected an original image from the set of images taken near the middle of the totality and performed  an absolute calibration of this image using well-known bright stars of the FOV following the method developed for photographic pictures taken at TSE using a neutral radial filter by \citet{Ko78}  and \citet{Le85}.  Some selected original images are available from the site \textsf{bass2000.obspm.fr/piwigo/index.php?/category/141}. We now have  digital 16-bit images composed from frames taken with different exposure times using a linear detector, which makes the work  easier. The details of this photometric analysis are beyond the scope of this work  and they will be given in an article in preparation. Absolute intensities of the K+F+S corona were deduced. K corresponds to the electron corona due to the Thomson scattering, F is the so-called dust corona due to the Mie scattering on particles orbiting the Sun and intercepted along the LOS, and S is the light due to the eclipse sky at totality and the eventual stray light easily evaluated using the intensity background measured on the disk of the Moon \citep{Ko78}. The excess intensities corresponding to the structure that we identified with the CME are found in the range of 3 to 6\,\% of the background intensities, and it is then easy to deduce the corresponding excess electron density [$n_{\rm e}^ {\rm CME}$] making the usual assumptions for computing the summed intensity along the LOS. 

We assumed that the effective thickness of the CME structure is typically 14 Mm (1/50\,R$_\odot$) and extends over 420 Mm (0.6\,R$_\odot$) at an average radial distance of 1.85\,R$_\odot$ where the effective length of integration along the LOS is 2\,R$_\odot$ \citep{No96}. Using our calibrated radial scans made at the latitude where the CME structure is identified, we deduced a background $n_{\rm e}$ density of $ 0.2 \times 10^7$\,cm$^{-3}$ at $r$ = 1.75\,R$_\odot$. The average density in the CME structure is then of order of $( 0.8 \pm 0.2) \times 10^7$\,cm$^{-3}$. This is a weighted average value taking into account both the radial variations and the azimuthal variations making the turbulent structure of the CME as caught on our TSE images. With these assumptions we deduced an overall mass of the eclipse CME at the SE (excess over the background local corona) of $1.6 \times 10^{12}$\,g. It seems to be the first evaluation of the mass of a CME made in the range of radial distances under 2\,R$_\odot$. We note that the CME catalogue assembled at NRL does not provide the mass of CMEs observed in 2017\,--\,18 as determined from the LASCO images; in any case they are the mass of CMEs at radial distances significantly greater than 2.5\,R$_\odot$ as observed after subtracting a minimum background resulting from a full coronal rotation. At a radial distance of 1.75\,R$_\odot$ the CME structure is made of pieces that are not all definitely showing an outward motion such that our estimate of mass corresponds to an upper limit \citep{Vo10}. It makes this CME a low mass CME (in the NRL terminology they would call it a poor CME) in addition to being a low-velocity one. Its morphology seems to suggest that it is the continuation of a process of outward flow of streamer sheets that occurs above the polar-crown filament and forms the shape of the classical streamer (see, {\it e.g.},  \citealp{Ko05}). The mass is clearly much lower than the mass of the erupted prominence.

\section{Discussion}

The study of CMEs is difficult due to the origin  of the event from a source region in the lower corona not observed in W-L in the FOV of the externally occulted space-based coronagraphs. SDO/AIA observes up to 0.3\,--\,0.5\,R$_\odot$  above the limb (depending on a position angle, because the detector is square) in EUV-lines reflecting different temperature regions. Images of the corona obtained using the Sun Watcher using Active Pixel System detector and Image Processing (SWAP) telescope onboard the {\it PRoject for Onboard Autonomy} (PROBA2: \citealp{Be06})  has a slightly larger FOV but  there is still a spatial gap between EUV observations and the inner boundary of SOHO/LASCO-C2 observations. In principle, the gap is covered by ground-based observations with the COronal Solar Magnetism Observatory (COSMO) K-coronagraph (K-Cor) at the Mauna Loa Solar Observatory (MLSO) in Hawaii having a FOV from 1.05 to 3\,R$_\odot$  but they have a rather limited spatial resolution ($\approx$ 6$^{\prime\prime}$) and, of course, daytime and weather limitations \citep{Ho13}. 

TSEs provide the unique opportunity to observe the corona from low heights with  good resolution. The observations are limited by the short period of totality at a specific ground-site and by the number of these sites along the path of the Moon shadow. Sometimes CMEs with conspicuous structure are seen in eclipse coronal images as, {\it e.g.}, during the 03  November 2013 eclipse. An eclipse CME was studied at the time of solar maximum \citep{Ko04}. However, it is usually difficult to recognize faint CMEs in W-L eclipse coronal images at solar minimum. Our new temporal sequence of high-resolution images taken during the totality in 2017 at a single site allows one to find rapid changes in the coronal structure and to identify a CME. Observations of this unique TSE and our image processing (\textsf{hdr-astrophotography.com/}) offered this opportunity for the first time.

Polar crown prominences, being stretched along heliographic parallels, can be viewed ``end on" at the limb. Coronal cavities surrounding polar-crown prominences are often observed in these locations because the LOS is directed along the cavity long axis. On 21 August 2017, a cavity manifested itself by a void in EUV emission and by U-shaped threads at the bottom of the cavity. It is widely understood that a cavity is the result of the helical magnetic-field lines of a flux rope. The lower parts of these field lines are observed as U-shaped structures in 171 \AA \ images. The lowest sections of the helices with dips may be filled with colder dense plasma of a prominence.  

Some instability of the flux rope causes it to erupt. We are not able to reach any conclusion about the type of instability (kink, torus, loss of equilibrium) based on the coronal magnetic field calculations because photospheric fields are weak below the polar-crown prominences and the surface of the photosphere make an acute angle with the LOS due to the closeness to the Pole and to the limb. Different photospheric changes can destabilize the flux rope. Usually, new magnetic-flux emergence \citep{St07,Ch13}, magnetic-flux cancellation \citep{Li99,Ya16}, photospheric plasma flows \citep{Ro08,Ro18}, and turbulence in the prominence \citep{Ko04} are considered as triggers of the eruption. We believe that the flux rope moves under the action of forces that controlled its equilibrium before the eruption, namely the Lorentz force, consisting of the holding part from the coronal magnetic field and the supporting part from the diamagnetic photosphere, and gravity \citep{va78,Mo87,Pr90}. Among them, only the diamagnetic force, which is often considered as interaction with the mirror image of the flux rope in the photosphere, is directed outward and accelerates plasma. The gravity can play a role in this process, since we deal with a rather slow process, and small cool blobs are observed returning to the surface .

Prominence material is also rising, but when the previously mostly straight  axis of the flux rope becomes arch-like, the dips become too shallow and the material can drain down to the chromosphere. The prominence seems to fail to erupt since  it fades out during the ascent before it reaches the FOV of the STEREO/EUVI. However, in contrast to many failed prominence eruptions, which are not followed by CMEs \citep{Fi02,Ji03,Li09,Gu10,Jo14}, the eruption on 21 August 2017 produced a faint, slow CME. \citet{Ku13} reported a slow CME observed after a prominence eruption, which failed or was too weak to be studied in detail. \citet{Ta18} analyzed a slow CME event that  also happened during a total solar eclipse. It was the 11 July 2010 eclipse. A small plasma blob or plasmoid was observed in EUV slowly accelerating up to a speed of 12\,km\,s$^{-1}$, then slowing down and stopped at a height of about 500 Mm in the corona. The blob was captured in W-L eclipse observations and the evaluated density of the blob is significantly bigger than what we evaluated in this event.  

Since the hot coronal structures were too faint to follow them near the upper boundary of the SDO/AIA FOV, it is not easy to decide whether the whole flux rope stopped in the middle corona as in two-step eruptions \citep{Fi02,By14,Go16,Wa16,Ch17}.  However, the observed mass-unloading  within the prominence arch favours  the flux rope scenario with acceleration as is suggested in mass-unloading eruption scenarios \citep{Kl01,Zh06,Je18}. If gravitation is significant in the flux-rope equilibrium, the acceleration after its violation cannot exceed the acceleration due to gravity [$g$  = 270\,m\,s$^{-2}$]. In such a process, only a rather slow CME can be produced, as  is the case in the 21 August 2017 event. The high-quality W-L coronal images obtained during the total eclipse prove that the erupting flux rope was present and moved in the corona after leaving the EUV telescope FOV and before the appearance in the coronagraph FOV. They allowed us to recognize a number of small coronal features moving outward in the south-east sector of the corona  during the short period of the totality and the presence of the prominence remnants at higher level than what was observed in the EUV.

\section{Summary and Conclusions}

  We analyzed the temporal changes within the SE part of the TSE solar corona before, during, and after the total solar eclipse on 21 August 2017. Short displacements of a number of coronal features were definitely found for the first time in W-L coronal images taken during the totality from observations made at a single observation site. Special image processing allows us to reveal these changes. They represent synchronous motion of the features away from the Sun with a speed of about 250\,km\,s$^{-1}$. The comparison with observations from space-based instruments onboard SDO, STEREO-A, and SOHO showed that the features belong to a slow CME propagating through the corona. SDO/AIA and STEREO-A/EUVI observations in the 304\,\AA \ channels showed a slow prominence eruption starting well before the totality. The prominence rose  with a speed of the order of 10\,km\,s$^{-1}$. The prominence material was observed to drain down to the chromosphere along the prominence arch preferentially to the south-west endpoint of the prominence. 

SDO/AIA 171\,\AA\ images show a coronal cavity around the initial position of the prominence. The lower edge of the cavity represented by the U-shaped structure ascended together with the prominence with a similar speed. We identify the cavity with the magnetic-flux rope containing the cool dense prominence material in dips at the bottom of presumably helical field lines. The flux rope lost its equilibrium and started to erupt due to some instability that we cannot specify on the basis of the available observations.

SOHO/LASCO data showed a slow CME propagating in the range from 3 to 15\,R$_\odot$  with a nearly constant speed of 250\,km\,s$^{-1}$. It  shows neither a prominent frontal loop nor a bright core. The start time of the CME is close to the beginning of the prominence activation. We assume this CME to be the continuation of the very slow prominence eruption. The draining down of the prominence material, possibly, unloaded the flux rope and allowed it to accelerate more energetically upward to form the slow CME. The slow CME speed and its low acceleration support this scenario because the driving force is comparable with the gravity force and the possible acceleration cannot exceed the gravity acceleration. The movie assembled during the short interval of totality of the eclipse observed at a single site with the same instrument (Movie 1) confirms this picture, suggesting that a whole set of the magnetic structures above the E-limb are indeed involved. Apparently, observations taken by different instruments along the path of totality did not provide a really useful comparison in order to improve the result of this analysis from a single-site observations taken with the excellent resolution. We found that it is important to use the same method and the same instrumental parameters to deduce relevant parameters. The forthcoming  ESA space mission PROBA 3  with its ASPIICS coronagraph \citep{La08,Re15,Sh19}  is well suited to tackle the questions that we left unresolved.

\begin{acks}[Acknowledgements]
 The authors thank the SOHO, the STEREO, and the SDO scientific teams for the high-quality data they supply. The CME catalog is generated and maintained at the CDAW Data Center by NASA and The Catholic University of America in cooperation with the Naval Research Laboratory. SOHO is a project of international cooperation between ESA and NASA. STEREO is the third mission in NASA�s Solar Terrestrial Probes program. SDO is a mission of NASA�s Living With a Star Program. Some movies were created using the ESA and NASA funded Helioviewer Project.
\end{acks}

\section*{Disclosure of Potential Conflicts of Interest}
The authors declare that they have no conflicts of interest.

\bibliographystyle{spr-mp-sola}
\bibliography{ref1}

\end{article} 

\end{document}